\documentclass[twocolumn,prl,showpacs,amsmath,amstex,amssymb,mathfonts]{revtex4-1}
\pdfoutput=1
\usepackage{amsthm,color,amsfonts,graphicx,verbatim}
\usepackage{amsmath}
\usepackage{amssymb}
\usepackage{amsthm}
\usepackage{amsfonts}
\usepackage{listings}
\usepackage{enumerate}
\usepackage{latexsym}
\usepackage{psfrag}
\usepackage{bm}
\usepackage{graphicx}
\newcommand{\beq}{\begin{equation}}
\newcommand{\eeq}{\end{equation}}
\newcommand{\beqn}{\begin{eqnarray}}
\newcommand{\eeqn}{\end{eqnarray}}

\usepackage{pdfpages}

\begin{document}
\title{Expansion potentials for exact far-from-equilibrium spreading of particles and energy}
\author{Romain Vasseur}
\affiliation{Department of Physics, University of California,
Berkeley, CA 94720}
\affiliation{Materials Sciences Division,
Lawrence Berkeley National Laboratory, Berkeley, CA 94720}
\author{Christoph Karrasch}
\affiliation{Department of Physics, University of California,
Berkeley, CA 94720}
\affiliation{Materials Sciences Division,
Lawrence Berkeley National Laboratory, Berkeley, CA 94720}
\author{Joel E. Moore}
\affiliation{Department of Physics, University of California,
Berkeley, CA 94720} \affiliation{Materials Sciences Division,
Lawrence Berkeley National Laboratory, Berkeley, CA 94720}

\date{\today}
\begin{abstract}
The rates at which energy and particle densities move to equalize arbitrarily large temperature and chemical potential differences in an isolated quantum system have an emergent thermodynamical description whenever energy or particle current commutes with the Hamiltonian.  Concrete examples include the energy current in the 1D spinless fermion model with nearest-neighbor interactions (XXZ spin chain), energy current in Lorentz-invariant theories or particle current in interacting Bose gases in arbitrary dimension.   Even far from equilibrium, these rates are controlled by state functions, which we call ``expansion potentials'', expressed as integrals of equilibrium Drude weights.  This relation between nonequilibrium quantities and linear response implies non-equilibrium Maxwell relations for the Drude weights.  We verify our results via DMRG calculations for the XXZ chain.
\end{abstract}
\maketitle

The dynamics of how a system of interacting particles expands from an initial state with spatial variation of temperature, density, or both is one of the basic problems in non-equilibrium statistical physics.  The study of quantum effects on this process was reinvigorated by the experimental creation of ultracold atomic gases~\cite{bloch_review,lamacraftmoore}, including cases where the atoms are confined to one or two spatial dimensions.  Originally the main quantity measured was the momentum distribution~\cite{bloch_tonksgirardeau,Kinoshita:2006p900}, but recent progress on the ``quantum gas microscope'' and related techniques has made it possible to image particle density with high resolution, e.g., on single sites of an optical lattice~\cite{greiner_gas,chin_gas,bloch_gas}.

Such imaging methods mean that important observables to characterize expansion of an atomic gas in either free space or an optical lattice~\cite{bloch_expansion} are not the same as those for non-equilibrium processes in electronic transport.  For electrons, the charge or energy current between two leads  has been studied in hundreds of situations, including a few non-equilibrium results with interactions such as tunneling between Luttinger liquids~\cite{KANE:1992tu,fendleyludwigsaleur}, the interacting resonant level model~\cite{PhysRevLett.99.076806,PhysRevB.77.033409,PhysRevLett.101.140601,PhysRevLett.112.216802} and the single impurity Anderson model (for a recent review see~\cite{1367-2630-12-4-043042}).  The point of the present work is to show that one natural quantity of interest for atomic expansion measurements~\cite{PhysRevA.66.021601,PhysRevLett.99.220601,PhysRevLett.101.155303,Hackermuller26032010,PhysRevLett.104.160403,Schneider:2012aa}, namely the change in time of the first moment of particle or energy density, has a precise non-equilibrium thermodynamic description in a broad class of systems.  
For a continuum system with either Lorentz or Galilean invariance, this description reduces to standard thermodynamic state functions, but we find that even lattice systems relevant to current experiments have a description in terms of an ``expansion potential''  that is distinct from conventional thermodynamic quantities.

We use this description to compute the energy expansion rate exactly in the anisotropic Heisenberg spin chain (XXZ model) and compare our results in detail against time-dependent density-matrix renormalization group (DMRG~\cite{whitetdmrg,vidal,schollwoeck}) calculations using the finite temperature algorithm explained in~\cite{1367-2630-15-8-083031}.  The same formalism is applicable to higher-dimensional systems with emergent Lorentz or Galilean invariance.  Our predictions apply in particular to a one-dimensional Bose gas (Lieb-Liniger model~\cite{lieb1963,yang:1967}, or its lattice regularization in terms of $q$-deformed bosons~\cite{0305-4470-25-14-020}), expanding into vacuum, a problem that has attracted a lot of attention recently~\cite{castin_dum,shlyapnikov_expansion,stringari_expansion,gangardt_pustilnik,fleischauer,caux_konik,PhysRevLett.98.050405,gritsev_demler,caux_essler,iyer_andrei,bloch_expansion,gangardtbose,PhysRevB.89.075139}. Our results show that, at least for some quantities, exact results can be obtained for far-from-equilibrium expansion even in lattice models at arbitrary coupling strength.

 \begin{figure}[b!]
\includegraphics[width=1.0\columnwidth]{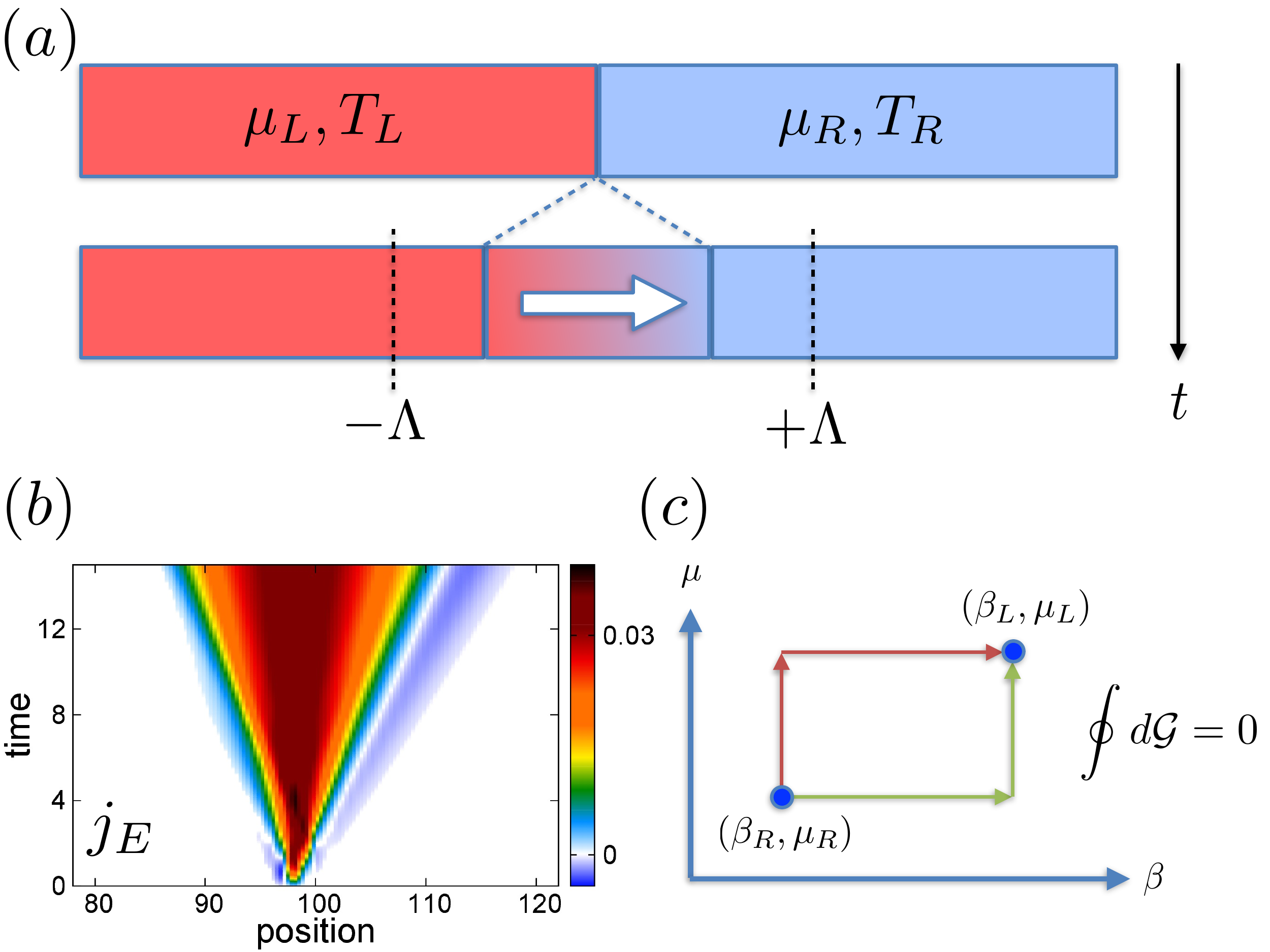}
\caption{(a) Nonequilibrium expansion setup considered in this letter. (b) Energy point current $j_E(x,t)$ in the XXZ spin chain (see Fig~\ref{Fig1} for parameters). (c) The variation of the expansion potential ${\cal G}$ does not depend on the path in the $(\beta,\mu)$ space. This implies nonequilibrium Maxwell relations (see text).}
\label{Fig0}
\end{figure}

At $t=0$, prepare two semi-infinite regions $x<0$ and $x>0$  at equilibrium with chemical potentials and temperatures $(\mu_L,T_L)$ and $(\mu_R,T_R)$ (Fig~\ref{Fig0}a).  (The initial state on the boundary between the two leads, or a possible finite extent of the boundary region, will not matter for the quantities of interest here after some initial transient.)  We write one-dimensional equations for simplicity but the concept is general.  We quantify the expansion for $t>0$ by the time dependence of the first moment of particle density, or similarly for energy,
\begin{equation}
M_1(t) = \int^{\Lambda}_{-\Lambda}\,n(x,t) x\,dx,
\end{equation}
with $\Lambda$ a large observation scale: $\Lambda \gg v t$ with $v$ a typical velocity.
From now on we suppress the arguments of $n$ and $M_1$.  The continuity equation relates density and current $\partial_t n + \partial_x j = 0$. Now
$\frac{d M_1}{dt} = - \int^{\Lambda}_{-\Lambda} \,x \partial_x j\,dx = {\cal J}$,
with ${\cal J}= \int^{\Lambda}_{-\Lambda} j\,dx$ where in the integration by parts we have assumed $j(x)\approx 0$ at $x = \pm \Lambda$ (Fig~\ref{Fig0}b).

 \begin{figure}[t!]
\includegraphics[width=1.0\columnwidth]{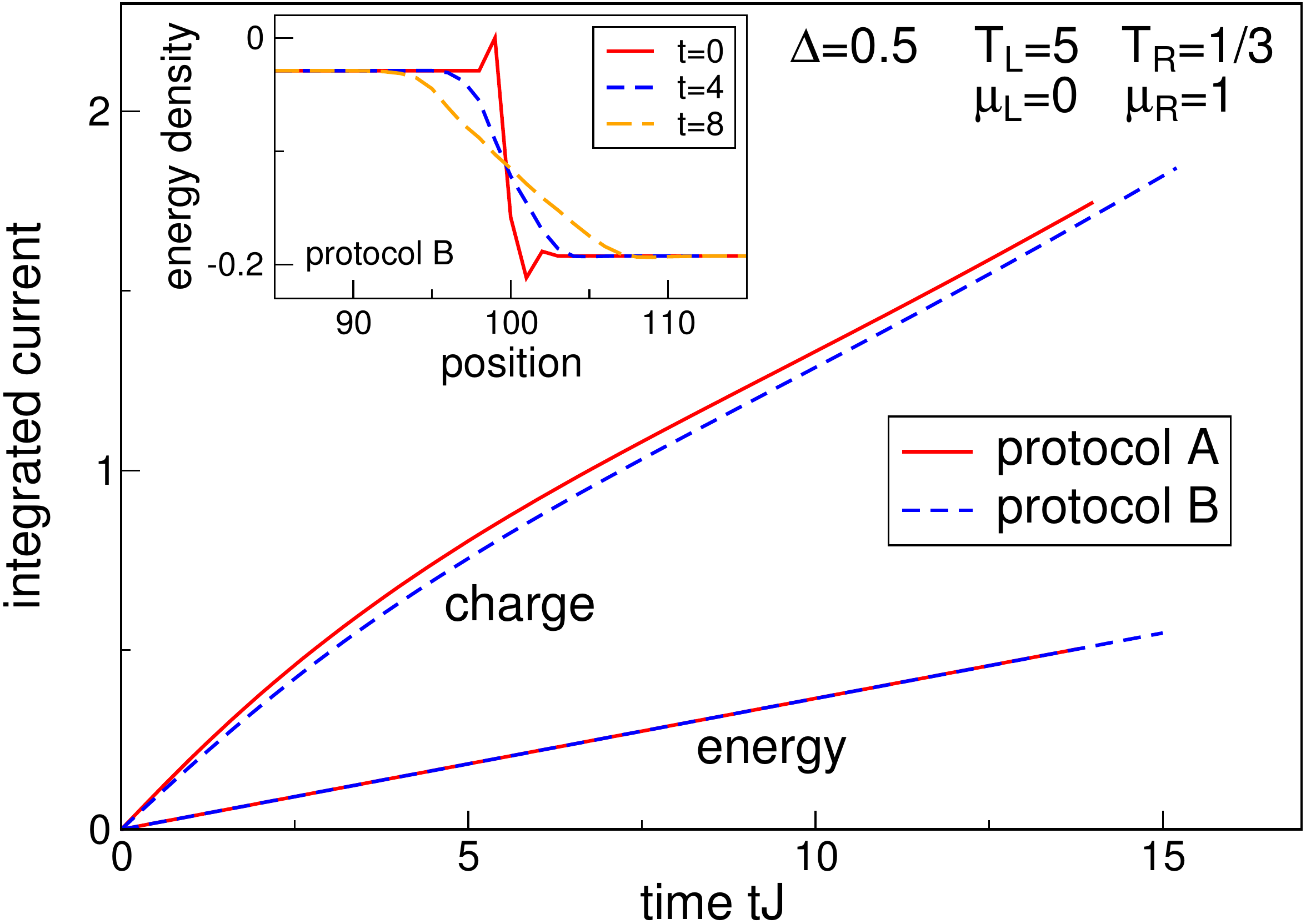}
\caption{Spatially integrated charge and energy currents in the XXZ model with open boundary conditions as a function of time for one choice of parameters $T_L$, $\mu_L$, $T_R$, $\mu_R$ at $\Delta=0.5$ under two protocols (A) and (B) that differ only in the way the central bond is dealt with in the initial state~\cite{SupMat}. The currents ${\cal J} = \int^\Lambda_{-\Lambda}  j dx$ are locally integrated around the cut site. The chemical potentials $\mu_{L,R}$ prepare the state but are not included in the real-time evolution (i.e., they are chemical potentials rather than electric potentials).  Inset: energy density profile as a function of time. The spatially integrated energy current is equal to $d M^{\rm th}_1/d t$ where $M^{\rm th}_1$ is the first moment of energy density.}
\label{Fig1}
\end{figure}

The key ingredient for the existence of an expansion potential is the conservation of integrated current:
\begin{equation}
\left[\oint j\,dx, H\right] = 0,
\label{currentcons}
\end{equation}
which is true for many problems of interest with periodic boundary conditions.
Note that this is a stronger statement than what is sometimes meant by a ``conserved current'', which is anything related to a conserved charge by a continuity equation.  A simple example with such a conservation law is a Bose gas in $d$ spatial dimensions with say, $\delta$-function interactions
%
$H  = \int d^dx \Psi^\dagger \left( - \frac{\nabla^2}{2m}  \right)\Psi + c \Psi^\dagger \Psi^\dagger \Psi  \Psi$,
%
with $\left[ \Psi^{\dagger}(x),\Psi(y) \right]=\delta(x-y)$, where the total particle current ${\cal J}_Q = - i \int dx (\Psi^\dagger \nabla \Psi - \nabla \Psi^\dagger \Psi)$ is conserved.  More generally, a system with one species of particles moving in the continuum in any spatial dimension will satisfy (\ref{currentcons}) for particle current if particle current is proportional to total momentum and momentum is conserved by the interactions.  A less trivial example of (\ref{currentcons}) is energy current in the spinless fermion model or XXZ spin chain (we will use the former representation): the energy current operator ${\cal J}_E = i \sum_j [h_j, h_{j+1}]$ commutes with the XXZ Hamiltonian $H^{\rm XXZ}  = \sum_i h_i$ with 
\beq
h_i= -\frac{J}{2}  \left(c_{i+1}^\dagger c_i + {\rm h.c.}  \right) + J \Delta \left(n_i-\frac{1}{2}\right)\left(n_{i+1}-\frac{1}{2}\right) ,
\eeq
with $n_i = c_i^\dagger c_i$, implying purely ballistic energy transport~\cite{Schwab:2000aa,SupMat}. An example of a current not conserved in this sense is {\it charge} current in the XXZ model; while there is a degree of ballistic transport in this model in the gapless regime even at nonzero temperature~\cite{prosen,karraschdrude,PhysRevB.87.245128}, the commutator in (\ref{currentcons}) is nonzero.
Steady-state energy currents between reservoirs have been actively studied~\cite{spyros,1751-8121-45-36-362001,karraschilanmoore,BernardDoyonReview,Doyon2015190,PhysRevB.90.161101,Bhaseen:2015aa} but exact results have been difficult to obtain except in the low-temperature conformal limit or for noninteracting systems.

\paragraph{Expansion potentials.}  

The global current conservation law~(\ref{currentcons}) implies that the current density should itself satisfy a continuity equation for some ``current of current'' $P$,
\beq
\partial_t j + \partial_x P = 0,
\label{secondcons}
\eeq
and we will see in the following that the operator $P$ is related to pressure for systems with emergent Galilean or Lorentz invariance.
Now spatially integrate this second continuity equation (\ref{secondcons}) over the region $[-\Lambda,\Lambda]$ centered on the boundary between our two large reservoirs $L$ and $R$.  Then
\beq
\frac{d^2 M_1}{dt^2}= - P \large]^\Lambda_{-\Lambda} = \Delta {\cal G}=  {\cal G}_L - {\cal G}_R,
\label{integratedcontinuity}
\eeq
where we have introduced the expansion potential ${\cal G}(\mu,T) =  \langle P \rangle_{\mu,T}$ for the thermodynamic expectation of the operator $P$. 

This is a strong constraint on the integrated current ${\cal J} = \int_{-\Lambda}^{\Lambda} j dx$.   We perform DMRG calculations on a XXZ spin chain with open boundary conditions that effectively describes a region of an infinite system. Within that region (shown in Fig.~\ref{Fig0}b), the total energy current is clearly not conserved~\cite{SupMat} and grows linearly with time (Fig.~\ref{Fig1}) for times  short enough that the reservoirs are effectively infinite, so their initial values can be used in the boundary evaluation on the right-hand side of~(\ref{integratedcontinuity}).
If the current has both diffusive and ballistic components (like the charge current in the XXZ chain), diffusive contributions die out after a transient and the spatially integrated current also grows linearly. However, the situation becomes especially simple for a current satisfying~\eqref{currentcons}, the key being that the right-hand side of~(\ref{integratedcontinuity}) contains only the operator $P$ evaluated {\it at equilibrium}, since deep within the reservoirs the system remains arbitrarily close to equilibrium in this intermediate time regime. This result relies only on~\eqref{currentcons} and does not depend on  whether the system is gapped or gapless for instance. 


 \begin{figure}[t!]
\includegraphics[width=1.0\columnwidth]{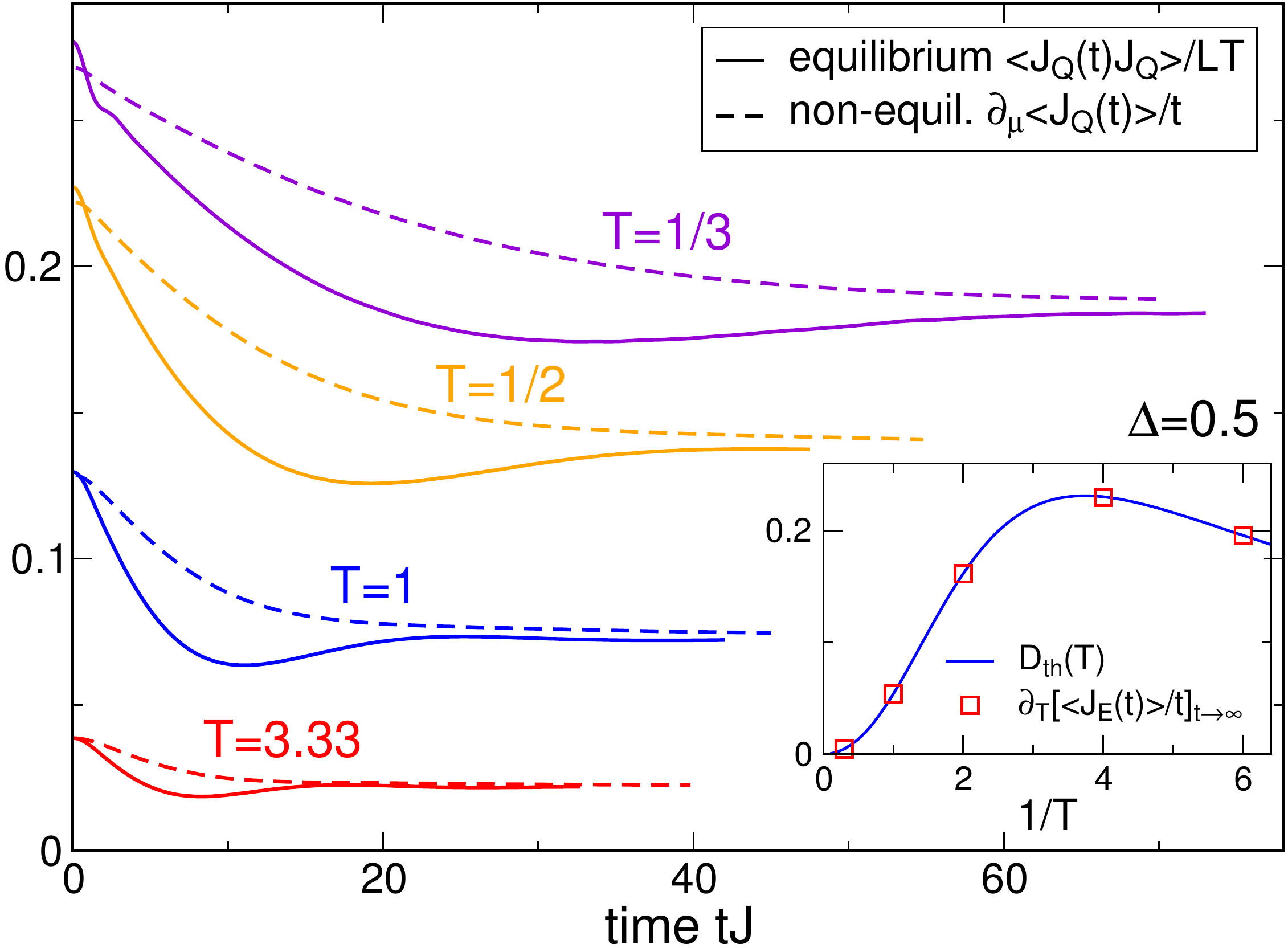}
\caption{Comparison between charge Drude weight (long-time asymptote of $\langle {\cal J}_Q(t) {\cal J}_Q(0)\rangle/LT$ and rate of particle spreading $ d^2 M^{\rm c}_1/dt^2 = \langle {\cal J}_Q(t)\rangle_t/t$ for a small chemical potential difference ($\Delta \mu \sim 10^{-3} \ll J=1$) in the XXZ chain (protocol A). Inset: similar relation for energy transport between thermal Drude weight and rate of energy spreading  $ d^2 M^{\rm th}_1/dt^2 = \langle {\cal J}_E(t)\rangle_t/t$ for a small temperature difference at half-filling (see eq.~\eqref{eqLinReponse}).}
\label{Fig2}
\end{figure}

\paragraph{Linear response.} One more relation is all that is needed to compute the expansion potential in some important cases. This is because eq.~\eqref{integratedcontinuity} implies that linear-response is enough to predict non-equilibrium, since linear response gives the derivative of ${\cal G}$, and knowing its derivative determines the function up to an arbitrary additive constant.  Focusing for the moment on energy current and a purely thermal gradient, linear response then predicts  $j_E = -\sigma_E \nabla T$ with the thermal conductivity characterized by a thermal Drude weight $ \sigma_E(\omega) = \pi D_{\rm th}(T) \delta(\omega)$ with $D_{\rm th} = \beta^2 \langle {\cal J}_E^2\rangle/L$, where $L$ is the size of the system. The spatially integrated current between the two reservoirs $R$ and $L$ then reads $ \int_{-\Lambda}^{\Lambda} j_E dx =  \pi \Delta T \delta(\omega=0) D_{\rm th}(T)$ where the time $t$ can be thought of as an infrared cutoff that regularizes $\delta(\omega=0) \approx \int^{t}_{-t} \frac{d t}{2 \pi} = t/ \pi$. We thus find
\begin{equation}
\frac{d^2 M^{\rm th}_1}{d t^2} = \frac{1}{t} \langle {\cal J}_E \rangle_t= D_{\rm th}(T)  \times (\Delta T),
\label{eqLinReponse}
\end{equation}
with ${\cal J}_E = \int_{-\Lambda}^{\Lambda} j_E dx$ and $\langle \dots \rangle_t$ refers to the nonequilibrium expectation value after time $t$. For the charge current at constant temperature $T_R = T_L =T$, we similarly find $\frac{d^2 M^{\rm c}_1}{d t^2} = \langle {\cal J}_Q\rangle_t /t = D_{c} \Delta \mu$ with the charge Drude weight $D_{c} = \beta \langle {\cal J}_Q^2\rangle/L $ (if $[H,{\cal J}_Q]=0$), for a small chemical potential gradient $\Delta \mu$. These results are easily extended to the case where both temperature and chemical potential gradients are present (see below). We also note that these linear response results remain valid even if the currents are not fully conserved and contain diffusive parts, like the charge current in the XXZ spin chain, which provides a direct way to measure Drude weights via imaging in cold atom experiments (see also~\cite{2013arXiv1306.4018H}). We checked this relation between charge ({\it resp.} thermal) Drude weight and linear-response rate of spreading of charge ({\it resp.} energy) in the XXZ chain (see Fig.~\ref{Fig2}) -- similar relations also exist for diffusive systems~\cite{hanggi}. 

\begin{figure}[t!]
\includegraphics[width=1.0\columnwidth]{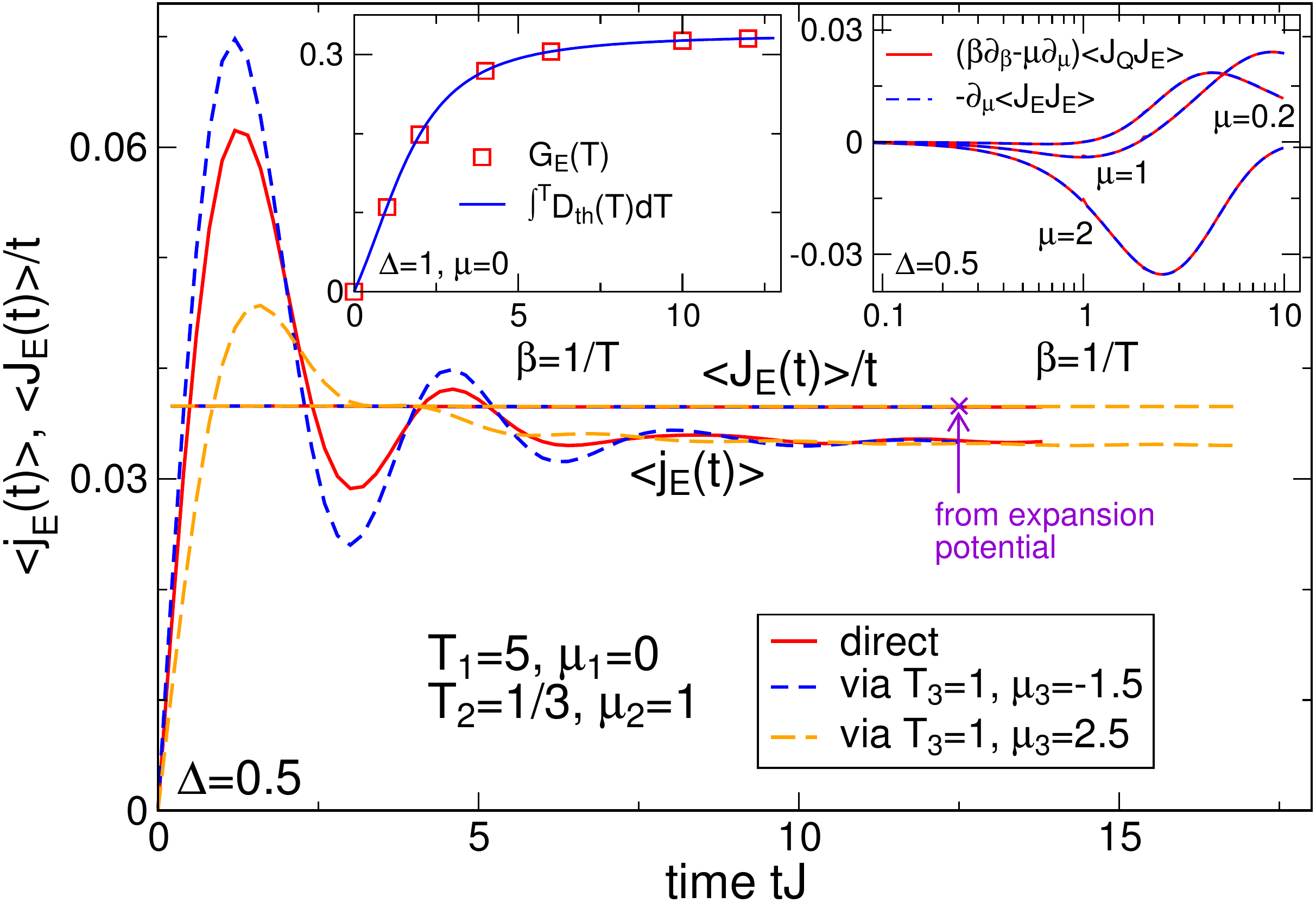}
\caption{The thermodynamic description~\eqref{integratedcontinuity} implies that the spatially integrated current ${\cal J}_{1\to 2}$ between two reservoirs $1$ and $2$ should be equal to ${\cal J}_{1\to 3}+{\cal J}_{3\to 2}$ for any intermediate reservoir 3. We verified this ``cyclic invariance'' of the spatially integrated energy current $\langle {\cal J}_E(t)\rangle_t/t$ and point current $\langle j_E(x=0,t)\rangle_t$ in the XXZ chain (protocol A). While cyclicity for the point current may be only approximate, it is exact for the integrated current. Insets: numerical check of eq.~\eqref{eqNoneqT} and of the nonequilibrium Maxwell relation~\eqref{eqMax1} that follows from conservation of energy current.}
\label{Fig3}
\end{figure}

\paragraph{Nonequilibrium expansion potentials.} The thermodynamic description eq.~\eqref{integratedcontinuity} together with the linear response prediction implies that the spreading of particles and energy far from equilibrium are fully characterized by the equilibrium Drude weights. As an example, let us consider the rate of energy spread in the XXZ spin chain between two reservoirs at different temperatures $T_R$ and $T_L$ and $\mu=0$. Then even far from equilibrium
\begin{equation}
\frac{d M^{\rm th}_1}{d t} \underset{t \to \infty}{\sim} t \times  \int_{T_R}^{T_L} D_{\rm th}(T) d T.
\label{eqNoneqT}
\end{equation}
In other words, the nonequilibrium rate of energy spread is given by the variation $\Delta_{R \to L} {\cal G}_E = {\cal G}_E(T_L)-{\cal G}_E(T_R)$ of a state function ${\cal G}_E(T)$ with $\partial_T {\cal G}_E = D_{\rm th}(T)$. This can be checked numerically by comparing the rate of expansion to the thermal Drude weight of the XXZ model computed by Kl\"umper and Sakai~\cite{0305-4470-35-9-307} (see Fig.~\ref{Fig3}).  

This is easily generalized to the case of reservoirs $R$ and $L$ with both different temperatures ($T_R$ and $T_L$) and chemical potentials ($\mu_R$ and $\mu_L$) . If the energy current is conserved, eq.~\eqref{integratedcontinuity} implies that the far-from-equilibrium rate of energy spread is given by the variation of an expansion potential ${\cal G}_E(\mu,\beta=T^{-1})$
\begin{equation}
\frac{d^2 M^{\rm th}_1}{d t^2} = \Delta_{R \to L} {\cal G}_E = \int_{R \to L} d {\cal G}_E , 
\end{equation}
where the differential $d {\cal G}_E$ is exact so that the integral does not depend on the chosen path. 
The state function ${\cal G}_E$ is then fully determined by the equilibrium Drude weights associated with the conservation of the energy current. Linear response theory~\cite{Mahan} then yields
\begin{equation}
d {\cal G}_E =\beta \frac{\langle {\cal J}_Q {\cal J}_E \rangle }{L} d\mu - \left( \frac{\langle {\cal J}_E^2 \rangle}{L} - \mu  \frac{\langle {\cal J}_Q {\cal J}_E \rangle}{L}\right) d \beta. 
\label{eqdGE}
\end{equation}
%
Even if ${\cal J}_Q$ is not conserved, the Drude thermopower is a thermodynamic quantity determined by $\langle {\cal J}_Q {\cal J}_E \rangle$ provided that $\left[{\cal J}_E,H\right]=0$. 
If the particle current is conserved, 
we find similarly that the integrated nonequilibrium particle current between two reservoirs $(\mu_R,\beta_R)$ and $(\mu_L,\beta_L)$ is given by the variation of another state function $\frac{d^2 M^{\rm c}_1}{d t^2}  = \Delta_{R \to L} {\cal G}_Q$ 
with 
$d {\cal G}_Q =\beta \frac{\langle {\cal J}_Q^2 \rangle }{L} d\mu - \left( \frac{\langle {\cal J}_Q {\cal J}_E \rangle}{L} - \mu  \frac{\langle {\cal J}_Q^2 \rangle}{L}\right) d \beta$. 

\paragraph{Nonequilibrium Maxwell relations.} We saw above that when either the energy or particle current is fully conserved, then even far from equilibrium the expansion dynamics of energy or particle  densities are characterized by state functions that are entirely determined by equilibrium Drude weights. An interesting corollary of the path-independence of these state functions (Fig~\ref{Fig0}c) are nonequilibrium Maxwell relations for the Drude weights. 
For example, if the energy current is conserved,  $\partial_\mu \partial_\beta {\cal G}_E = \partial_\beta \partial_\mu {\cal G}_E$ yields 
%
\begin{equation}
(\beta \partial_\beta - \mu \partial_\mu) \langle {\cal J}_Q {\cal J}_E \rangle + \partial_\mu \langle {\cal J}_E^2 \rangle = 0, 
\label{eqMax1}
\end{equation}
which can also be rewritten as $\langle \Delta H {\cal J}_Q {\cal J}_{E} \rangle=\langle \Delta N {\cal J}_E^2\rangle$ with $\Delta H = H -\langle H \rangle$ and $\Delta N = N -\langle N \rangle$. This equality was known in the context of the XXZ chain~\cite{PhysRevB.67.224410} and was actually used to compute the Drude thermopower analytically~\cite{doi:10.1143/JPSJS.74S.196}, but our approach provides a very transparent derivation of why such a relation has to hold (see Fig.~\ref{Fig3} for a numerical check).  
If the charge current is conserved, then the associated nonequilibrium Maxwell relation reads 
$(\beta \partial_\beta - \mu \partial_\mu) \langle {\cal J}_Q^2 \rangle + \partial_\mu \langle {\cal J}_E {\cal J}_Q \rangle = 0$,
which can also be rewritten as $\langle \Delta H {\cal J}_Q^2 \rangle=\langle \Delta N {\cal J}_Q {\cal J}_E \rangle$.

\paragraph{Examples in $d>1$ dimensions.} Even though most of the arguments discussed above focused on one dimension for simplicity, the general concepts apply in higher dimension as well. For a system with emergent Lorentz symmetry ($z=1$ critical points for instance), the symmetry of the stress-energy tensor means that the energy current $T_{0i}$ with $i=1,\dots,d$ is also the (conserved) momentum density $T_{i0}$. The energy expansion potential then reads ${\cal G}_E(\beta) = - \int^\beta\,d\beta \int d^d x \frac{1}{d}\sum_i \langle T_{0i}(x)T_{0i}(0) \rangle$, which can be related to pressure~\cite{Doyon2015190,Bhaseen:2015aa}. In a non-relativistic system with a single species of particles and current proportional to (conserved) momentum, there is a particle expansion potential; the interacting Bose gas is one such example. The particle Drude weight $D_c$ is then entirely determined by the sum rule $\int \frac{d \omega}{\pi} \sigma(\omega) = D_c = \frac{n}{m}$ with $n$ the density and $m$ the mass~\cite{Mahan}. This immediately implies that the expansion potential is simply related to pressure ${\cal G}_Q = -\frac{\Omega}{V m} = \frac{P}{m}$ with $\Omega$ the thermodynamic grand potential and $V$ the volume -- this is a consequence of Galilean invariance~\footnote{A technical condition for direct application of our results is that the reservoirs be uniform, rather than having a parabolic confining potential, but a precisely characterized uniform atomic trap has recently been achieved and used to probe expansion~\cite{hadzibabic_uniform,hadzibabic2}.
}. The Drude thermopower is then given by $\langle {\cal J}_Q {\cal J}_E \rangle/V = \frac{T}{m} (u+ P)$ with $u$ the internal energy density. These quantities can be computed explicitly for the Lieb-Liniger gas in one dimension as a function of $T$ and $\mu$ (or particle density)~\cite{YangYang}.  This and other simple cases where the expansion potentials can be computed explicitly, such as non-interacting systems and Luttinger liquids, are given in Supplemental Material~\cite{SupMat}.

\paragraph{Nature of the steady-state.} Interestingly, the variation of expansion potential $\Delta {\cal G}$ provides a lower bound for the point current $j(x)$~\cite{Doyon2015190}. However, the more general relation between spatially integrated and point currents remains mysterious. We find numerically that both the energy density $n_E(x,t)$ and the energy current $j_E(x,t)$ in the XXZ spin chain at half-filling become functions of $x/t$ at large enough times, with nontrivial limiting shapes~\cite{SupMat}. In the low-temperature limit described by conformal field theory~\cite{1751-8121-45-36-362001,BernardDoyonReview}, we expect a uniform steady-state local current $j_E(x)= \frac{\Delta {\cal G}}{2v}=\frac{\pi}{12} (T_L^2-T_R^2)$ over a region of size $2 v t$ with $v$ the spinon velocity. However, we find that the rescaled functions $j_E(x/t)$, $ n_E(x/t)$ even at moderate temperatures are very far from that picture: in general, there is no nonzero range of the reduced variable $x/t$ for which $j_E(x/t$) is constant, indicating that the steady-state region spreads sub-ballistically, and there are no transient ``shock-waves'' like those expected in the presence of Lorentz invariance~\cite{Bhaseen:2015aa,bernarddoyonunpub} separating the uniform steady-state region from the reservoirs.  It is an interesting problem for future work to determine more properties of the limiting function $j_E(x/t)$, possibly by adapting the recently developed hydrodynamic approaches for relativistic systems~\cite{Bhaseen:2015aa,bernarddoyonunpub} to incorporate the additional conserved quantities of integrable lattice spin chains.

\paragraph{Discussion.} 

In closing, we emphasize that the expansion potentials generalize familiar concepts in the presence of either Galilean or Lorentz invariance to considerably more complex physical situations.  Lattice models for which a current is conserved in the sense of~(\ref{currentcons}) include the XYZ spin chain, the $q$-Bose gas~\cite{0305-4470-25-14-020}, and the supersymmetric point of the $t$-$J$ model~\cite{PhysRevB.46.9147}. For systems where the conservation law does not strictly hold, such as the Bose-Hubbard model at small occupancy where rare double occupancies spoil the mapping to the XXZ model, Joule heating and other strongly non-equilibrium physics could be computed using perturbation theory from the expansion-potential case.  It would be interesting to connect the expansion potential to other nonequilibrium effects, such as ``quantum quenches'' of a coupling~\cite{Calabrese:2006}, which can reveal topological phases~\cite{sengupta2008exact,vasseurdahlhausmoore}. For lattice models with conserved energy current but without full integrability, the expansion potential still exists and could be computed numerically {\it at equilibrium}, while it would serve as a useful constraint on predictions about far-from-equilbrium energy flow~\cite{bernarddoyonunpub}.

\smallskip

\paragraph{Acknowledgments.}
The authors thank M.J. Bhaseen, B. Doyon, F. Essler, S. Gazit, V. Korepin, A.C. Potter, D. Weld, the Department of Energy through programs Thermoelectrics (C.K.) and Quantum Materials (R.V.), NSF DMR-1206535 and a Simons Investigatorship (J.E.M.), and center support from CaIQuE and the Moore Foundation's EPiQS initiative.

\bibliography{energy_transport}

\bigskip

\onecolumngrid

\newpage



\section*{Supplementary material}

# Supplemental Material for "Expansion potentials for exact far-from-equilibrium spreading of particles and energy"

Romain Vasseur, Christoph Karrasch, and Joel E. Moore
*Department of Physics, University of California, Berkeley, CA 94720 and*
*Materials Sciences Division, Lawrence Berkeley National Laboratory, Berkeley, CA 94720*

## NUMERICAL METHODS AND BOUNDARY CONDITIONS

We run our DMRG calculations using the following numerical parameters. The system size is typically chosen as $L = 200$, and for a few cases we check our results against $L = 100$ and $L = 300$ to ensure that we are effectively in the thermodynamic limit. The imaginary time evolution from $T = \infty$ down to the desired temperature is carried out using a second order Trotter decomposition with a step size of $\delta\tau = 0.01$, and the discarded weight during each individual sub-step is at most $10^{-12}$. The real time evolution is performed using a fourth order Trotter decomposition with a step size of $\delta t = 0.2$. The calculations are performed for various values of the discarded weight, which eventually is chosen such that physical quantities have a relative accuracy of approximately 1 percent. Typical values are $10^{-7\ldots-9}$ for the linear-reponse correlators as well as $10^{-9\ldots-11}$ for the non-equilibrium currents at small bias in Fig. 2 or $10^{-6}$ for the currents in presence of a large bias in Fig. 3. A simple test of the method is that the free case is reproduced exactly and the results do not change on increasing the size of the matrix product states used in the calculation.

We use numerically two protocols A and B to prepare the XXZ spin chain out of equilibrium. For protocol A, the system is prepared with the density matrix $e^{-\beta_L(\tilde{H}_L - \mu_L \tilde{N}_L) - \beta_R(H_R - \mu_R N_R)}$, where $\tilde{H}_L$ and $\tilde{N}_L$ are the Hamiltonian and number operator in the left system including the bond to the right system. The chemical potential at the interface (i.e., at the first site of the right system) is chosen as $(\mu_L + \mu_R)/2$. For protocol B, this central bond is initially cut, and the density matrix reads $e^{-\beta_L(H_L - \mu_L N_L)} \otimes e^{-\beta_R(H_R - \mu_R N_R)}$. The chemical potentials $\mu_{L,R}$ prepare the state but are not included in the real-time evolution (i.e., they are chemical potentials rather than electric potentials). We find that both protocols give the same long-time limit for the spatially integrated currents $\mathcal{J}_E/t$ and $\mathcal{J}_Q/t$.

We now comment on the boundary conditions used in our DMRG calculation, and explain how an exact conservation law is consistent with the change in the integrated current observed in our numerics. As discussed in the main text, the conservation law $[H, \mathcal{J}_E] = 0$ for the XXZ spin chain is valid only for a system with periodic boundary conditions, with $\mathcal{J}_E = \oint j_E dx$. For such a periodic chain with two halves $A$ and $B$ prepared at different temperatures $T_A$ and $T_B$, our main results apply to the partially integrated current $\int_{-\Lambda}^{\Lambda} j_E dx = t \times (\mathcal{G}_A - \mathcal{G}_B)$ centered over one of the two junction regions between $A$ and $B$. Note that this does not contradict the conservation of the total spatially integrated current, as there is a net partially integrated current moving in the opposite direction at the opposite junction, so that the total current $\oint j_E dx = t \times (\mathcal{G}_A - \mathcal{G}_B) + t \times (\mathcal{G}_B - \mathcal{G}_A) = 0$ as required by the conservation law (Fig. 1). Our DMRG method is applied to a XXZ spin chain with open boundary conditions that effectively describes a region of an infinite system, and clearly within that region the total current is not conserved, but there are non-trivial consequences of the local current conservation $\partial_t j_E + \partial_x P_E = 0$ as described in the main text.

## BALLISTIC TRANSPORT

In this appendix, we provide a brief review of ballistic heat transport for the sake of completeness. Perhaps the most convincing experiment showing an extreme violation of Fourier's law for heat transport is the observation of the "thermal conductance quantum" [1]. We will review this famous 1D example of how heat in certain systems is characterized by ballistic transport and a quantized *conductance* rather than a finite *conductivity* a la Fourier. The XXZ model analyzed in the main text, which appears frequently in atomic physics, is an example of an interacting model that, because of its exact integrability, shows ballistic properties in its heat conduction, but it seems worthwhile to first discuss the free case before moving to the non-trivial interacting case.

What is the thermal conductance of a free gas of one-dimensional particles, e.g., bosons such as the phonon? One might think it is infinite as they propagate without scattering, but as for electrons it pays to be careful. The charge conductance of a 1D gas of electrons at low temperature was shown to be $G = e^2/h$ "per channel", i.e., per Fermi

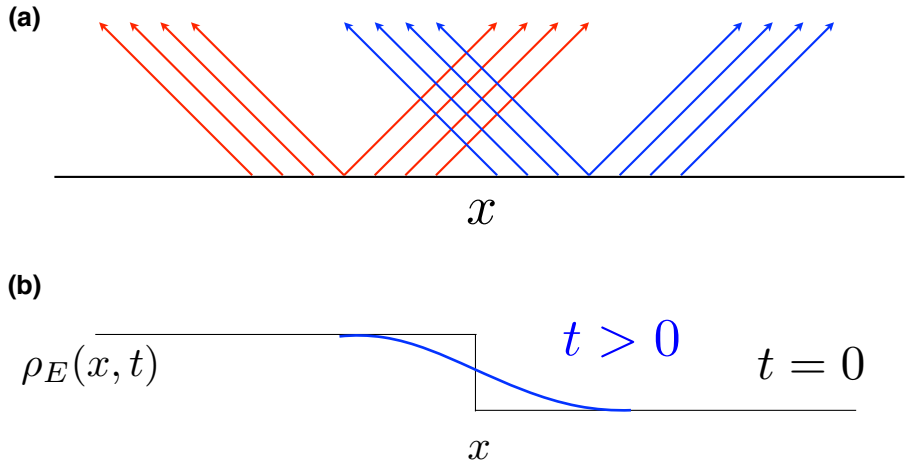

FIG. 1: (a) Energy transport of free particles: the left and right halves independently radiate at different temperatures, and the difference in radiated energy determines the thermal current. (b) Segment of an infinite 1D system with left and right halves initially prepared at different temperatures. Energy density evolves as shown in the region of the boundary, which means that there is a net rightward current of energy. This obviously does not violate conservation of spatial integral of energy current–the same local increase would hold for a segment of a large periodic system until such times as the two junction regions between high and low temperatures notice each other. In an infinite system, that time diverges.

surface point, rather than being infinite. What about the energy current if the two leads are at different temperatures instead of different voltages?

We can use the Landauer approach to compute this: for the case of perfect transmission, this amounts to computing the right-moving energy current from a lead at temperature $T + dt$ and subtracting the left-moving energy current from a lead at $T$. Assume one-dimensional free bosons as in the Schwab et al. experiment mentioned above. Using $k$ for momentum, we have that the total energy current (units of energy per time) is

$$J_E = J_E^R - J_E^L = \int_0^{\pi/a} \frac{dk}{2\pi} \left[ f_{T+dt}(\hbar\omega_k) - f_T(\hbar\omega_k) \right] \hbar\omega_k v_k. \tag{1}$$

Here $v_k = d\omega_k/dk$ and $f_T(E)$ is the Bose factor $(e^{E/k_B T} - 1)^{-1}$. So

$$J_E = (dt) \int_0^{\omega_{\max}} \frac{d\omega}{2\pi k_B T^2} \frac{e^{\hbar\omega/k_B T}}{(e^{\hbar\omega/k_B T} - 1)^2} \hbar^2 \omega^2. \tag{2}$$

Here $\omega_{\max}$ is the highest phonon frequency. If we assume that the temperature is small compared to this, so that $x = \hbar\omega/k_B T$ runs from 0 to infinity, then we obtain (note that we need to multiply by $(k_B T/\hbar)^3$)

$$J_E = \frac{k_B{}^2 T}{2\pi\hbar} (dt) \int_0^\infty dx \, \frac{x^2 e^x}{(e^x - 1)^2}. \tag{3}$$

The dimensionless integral gives $\pi^2/3$, so

$$G_0 = \frac{J_E}{dt} = \frac{\pi^2 k_B{}^2 T}{3h}. \tag{4}$$

An interesting fact about the thermal conductance $G_0$ is that it is the same for bosons or fermions (or indeed anyons), unlike charge transport. The Schwab et al. experiment observed one thermal conductance quantum $G_0$ for each low-temperature phonon mode.

Ballistic heat transport consistent with (4), with a factor of central charge, was recovered in the low-temperature limit of any 1D CFT by Sotiriadis and Cardy [2] and by Bernard and Doyon [3]. This result can be viewed as a form of the Stefan-Boltzmann law (see Fig. 1a): when particles are moving freely, the left half-line and right half-line both radiate particles that move directly through each other, so the thermal current is related to the difference in the Stefan-Boltzmann law of total radiated power between the left and right halves.

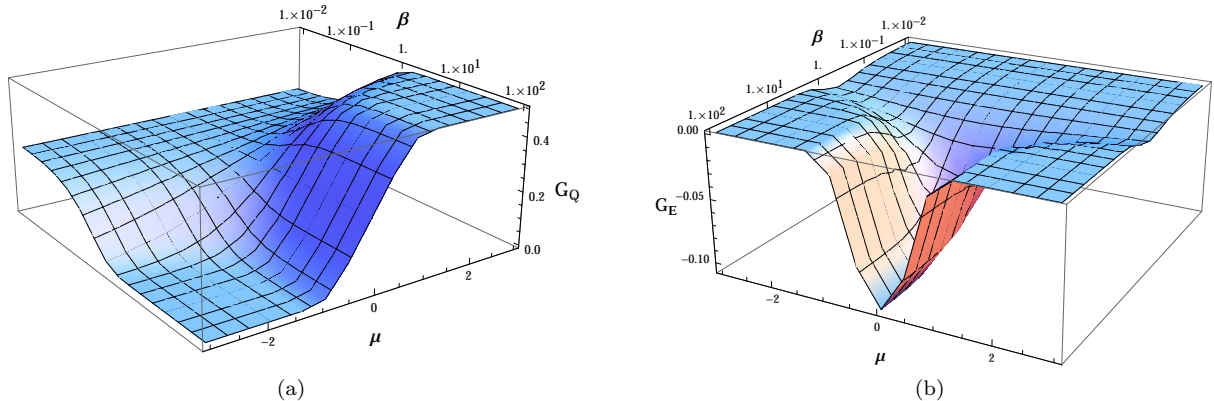

FIG. 2: Expansion potentials $\mathcal{G}_Q(\mu,\beta)$ (a) and $\mathcal{G}_E(\mu,\beta)$ (b) characterizing the spatially integrated nonequilibrium spin and energy currents in the XX chain.

## EXPANSION POTENTIALS FOR NON-INTERACTING FERMIONS

In this appendix we compute the energy and spin/charge expansion potentials $\mathcal{G}_Q(\beta,\mu)$ and $\mathcal{G}_E(\beta,\mu)$ in the XX chain (XXZ at $\Delta=0$)

$$H_{\text{XX}} = J \sum_i \left( S_i^x S_{i+1}^x + S_i^y S_{i+1}^y \right), \tag{5}$$

which can be mapped onto non-interacting fermions with dispersion $\epsilon_k = -\cos k$ — here and in the following, we will set $J=1$. Both the charge current $\mathcal{J}_Q = \sum_k v_k n_k$ and energy current $\mathcal{J}_E = \sum_k v_k n_k \epsilon_k$ commute with the Hamiltonian $H_{\text{XX}} = \sum_k \epsilon_k n_k$, where $n_k = c_k^\dagger c_k$ is the occupation number of the level $k$ and $v_k = \partial_k \epsilon_k = \sin k$. Therefore, as explained in the main text, the nonequilibrium spreading of particle and energy are fully characterized by the equilibrium Drude weights $\langle \mathcal{J}_E^2 \rangle$, $\langle \mathcal{J}_Q^2 \rangle$ and $\langle \mathcal{J}_E \mathcal{J}_Q \rangle$. Those can readily be computed from the generating function

$$g(\beta,\mu,\lambda_Q,\lambda_E) \equiv -\frac{T}{L} \ln \text{Tr } e^{-\beta(H-\mu N) + \lambda_Q \mathcal{J}_Q + \lambda_E \mathcal{J}_E}. \tag{6}$$

It is then straightforward to verify that they satisfy the nonequilibrium Maxwell relations $(\beta\partial_\beta - \mu\partial_\mu)\langle \mathcal{J}_Q \mathcal{J}_E \rangle + \partial_\mu \langle \mathcal{J}_E^2 \rangle = 0$ and $(\beta\partial_\beta - \mu\partial_\mu)\langle \mathcal{J}_Q^2 \rangle + \partial_\mu \langle \mathcal{J}_E \mathcal{J}_Q \rangle = 0$ as they should. We also remark that since both charge and energy currents are conserved, the charge and energy grand expansion potentials are related through $\partial_\mu \mathcal{G}_E = (\mu\partial_\mu - \beta\partial_\beta)\mathcal{G}_Q$. The expansion potentials are then obtained by integrating the differentials $d\mathcal{G}_E = \beta\frac{\langle \mathcal{J}_Q \mathcal{J}_E \rangle}{L} d\mu - \left( \frac{\langle \mathcal{J}_E^2 \rangle}{L} - \mu \frac{\langle \mathcal{J}_Q \mathcal{J}_E \rangle}{L} \right) d\beta$ and $d\mathcal{G}_Q = \beta\frac{\langle \mathcal{J}_Q^2 \rangle}{L} d\mu - \left( \frac{\langle \mathcal{J}_Q \mathcal{J}_E \rangle}{L} - \mu \frac{\langle \mathcal{J}_Q^2 \rangle}{L} \right) d\beta$:

$$\mathcal{G}_Q(\beta,\mu) = \int_{-\pi}^{\pi} \frac{dk}{2\pi} \frac{v_k^2}{1 + e^{\beta(\epsilon_k - \mu)}}, \tag{7}$$

and

$$\mathcal{G}_E(\beta,\mu) = \int_{-\pi}^{\pi} \frac{dk}{2\pi} \frac{v_k^2 \epsilon_k}{1 + e^{\beta(\epsilon_k - \mu)}}. \tag{8}$$

Focusing on energy transport, $\mathcal{G}_E(\beta,\mu) = -\frac{1}{3\pi} + \frac{\pi}{6} T^2 + \ldots$ at half-filling and low temperature. We remark that the $T^2$ term is entirely fixed by conformal field theory: it coincides with $2 v_F g(T)$ where $v_F = 1$ is the Fermi velocity and $g(T) = \frac{\pi}{12} T^2$ is the function characterizing the energy point current $j_E = g(T_L) - g(T_R)$ in the steady-state region [3] (and we have used the fact that the central charge $c=1$ for the XX chain). This is consistent with a steady-state region expanding as $[-v_F t, v_F t]$ with uniform point-current $j_E = g(T_L) - g(T_R)$, leading to the spatially integrated current $\int j_E dx = 2 v_F t (g(T_L) - g(T_R)) = t(\mathcal{G}(T_L) - \mathcal{G}(T_R))$. We emphasize however that this relation between point-current and expansion potential holds only in the low-energy regime. Whereas we have argued that the far-from-equilibrium spatially integrated current can be computed (even at high temperature) in terms of an expansion potential fully characterized by equilibrium Drude weights, it remains unclear whether a similar relation also holds for the point current.

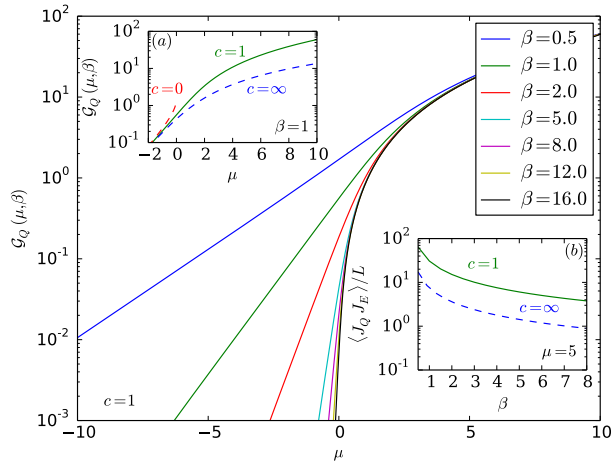

FIG. 3: Grand expansion potential $\mathcal{G}_Q(\mu, \beta) = \frac{P}{m}$ characterizing the non-equilibrium rate of particle density spread in the 1D Bose gas with $\delta$-function interactions with interaction strength $c = 1$. Inset (a): influence of the interaction strength on $\mathcal{G}_Q(\mu, \beta)$ and comparison with the $c = 0$ (non-interacting Bose gas) and $c = \infty$ (effectively free-fermions) limits. Inset (b): Thermopower Drude weight determined from the nonequilibrium Maxwell relation.

## EXPANSION POTENTIALS FOR A LUTTINGER LIQUID

In this appendix, we derive the form of the particle and energy expansion potentials for a Luttinger liquid. The low-energy degrees of freedom of the XXZ chain in the massless regime can be described by the Luttinger Hamiltonian

$$H = \frac{v}{2} \int dx \left( K(\partial_x \theta)^2 + K^{-1}(\partial_x \phi)^2 \right), \tag{9}$$

where the velocity $v$ and the Luttinger parameter $K$ depend in a non-universal way on microscopic parameters. The chemical potential $\mu$ is then introduced by coupling to the particle number $N = \int dx \partial_x \phi / \sqrt{\pi}$ – note that we consider $v$ and $K$ to be constants in the field theory description, even though they depend on the chemical potential/Fermi velocity of the underlying microscopic model. The particle and energy currents read $\mathcal{J}_Q = vK \int dx \partial_x \theta / \sqrt{\pi}$ and $\mathcal{J}_E = v^2 \int dx \partial_x \theta \partial_x \phi$; they both commute with the low-energy Hamiltonian $H$. The corresponding Drude weights are easily computed [4, 5] – note that in particular, $\langle (\mathcal{J}_E - \mu \mathcal{J}_Q) \mathcal{J}_Q \rangle = 0$. This yields

$$d\mathcal{G}_Q = \frac{vK}{\pi} d\mu, \tag{10}$$

$$d\mathcal{G}_E = \frac{\mu vK}{\pi} d\mu - \frac{\pi v}{3\beta^3} d\beta. \tag{11}$$

Within this low-energy limit, the expansion potentials are simply related to the value of the steady-state local (point) currents $j_Q = \frac{K}{2\pi}(\mu_L - \mu_R)$ and $j_E = \frac{\pi}{12}(T_L^2 - T_R^2) + \frac{K}{4\pi}(\mu_L^2 - \mu_R^2)$ by a factor $2v$ (the size of the steady-state region being $\sim 2vt$).

## PARTICLE DENSITY EXPANSION IN THE INTERACTING BOSE GAS

In this appendix, we discuss the particle expansion dynamics in the interacting Bose gas with $\delta$-function interactions. We start with the one-dimensional (integrable) Lieb-Liniger model

$$H = -\sum_i \frac{\partial^2}{\partial x_i^2} + 2c \sum_{i<j} \delta(x_i - x_j). \tag{12}$$

The Lieb-Liniger model was realized experimentally in various circumstances [6–9]. Since the particle current is conserved, the far-from-equilibrium expansion dynamics is entirely determined by the state function $\mathcal{G}_Q$. The

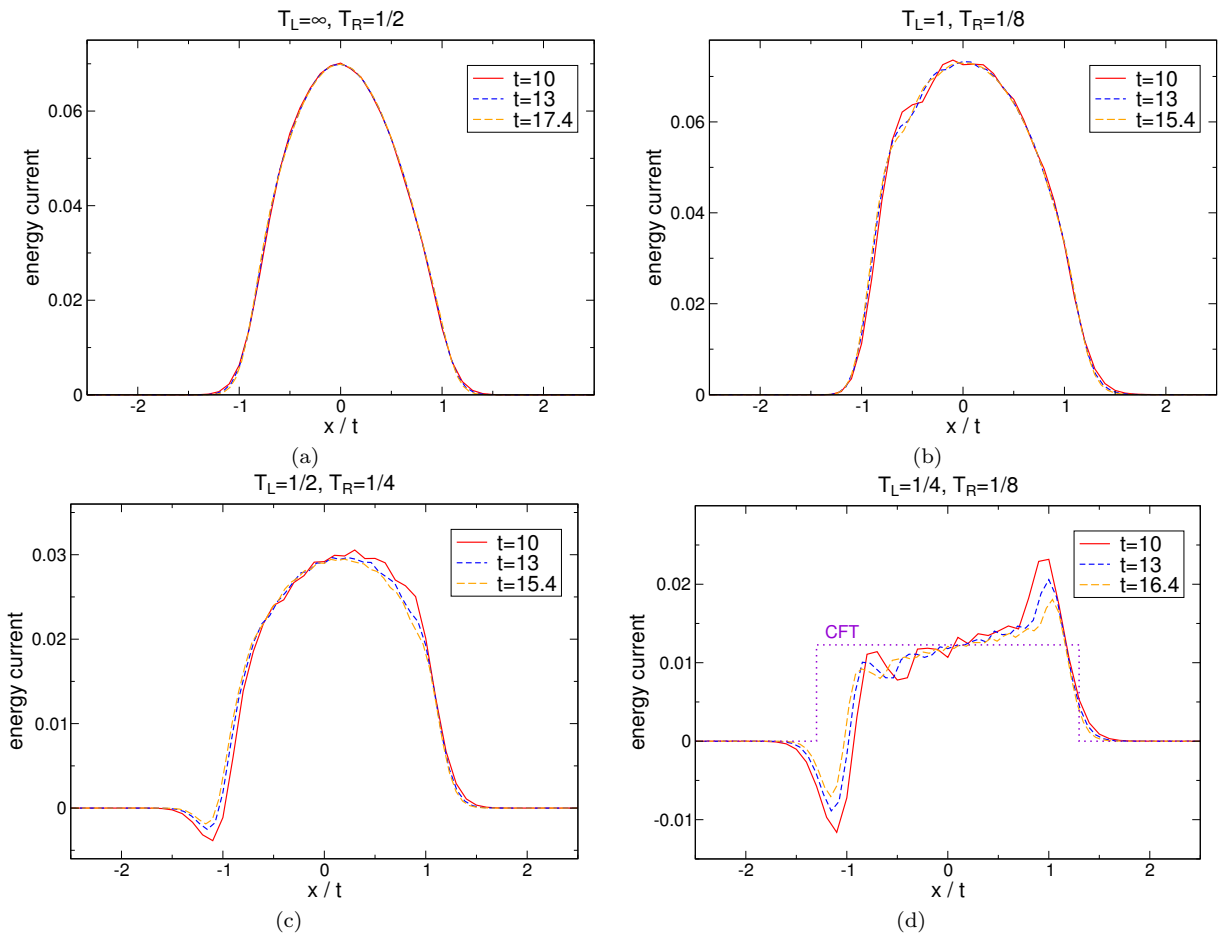

FIG. 4: Energy point current $j_E(x,t)$ as a function of $x/t$ in the XXZ spin chain at half-filling for various values of the parameters $T_L$ and $T_R$.

Drude weight $\langle \mathcal{J}_Q^2 \rangle/L$ can be obtained from the generating function $-\beta g(\beta,\mu,\lambda) = \frac{1}{L} \ln \left( \text{Tr } e^{-\beta(H-\mu N)+\lambda \mathcal{J}_Q} \right)$ as $\langle \mathcal{J}_Q^2 \rangle/L = \partial_\lambda^2 (-\beta g)|_{\lambda=0}$. Since the Lieb-Liniger model is integrable, the function $g(\beta,\mu,\lambda)$ can be computed using the Thermodynamic Bethe Ansatz (TBA) [10, 11] as $g(\beta,\mu,\lambda) = -T \int \frac{dk}{2\pi} \ln(1+e^{-\epsilon(k)/T})$ with $\epsilon(k)$ the pseudo-energy dressed by the interactions that can be determined from the non-linear integral equation

$$\epsilon(k)/T = \beta(k^2-\mu) - 2\lambda k - \frac{1}{\pi} \int \frac{c}{c^2+(k-q)^2} \ln\left(1+e^{-\epsilon(k)/T}\right) dq. \tag{13}$$

The nonequilibrium Maxwell relation $(\beta \partial_\beta - \mu \partial_\mu)\langle \mathcal{J}_Q^2 \rangle + \partial_\mu \langle \mathcal{J}_E \mathcal{J}_Q \rangle = 0$ then gives the Drude thermopower $\langle \mathcal{J}_E \mathcal{J}_Q \rangle/L$, which is an interesting quantity since thermoelectric properties can be measured in atomic gases [12], and the nonequilibrium Maxwell relation is crucial to compute it. Here we are working at fixed chemical potential rather than number density, but note that as the expansion process takes place in a closed quantum system and the initial reservoirs are thermodynamically large, the expansion rate for a density difference follows from working at the chemical potentials corresponding to the initial densities.

As mentioned in the main text, the particle Drude weight can also be obtained much more easily from the sum rule $\int \frac{d\omega}{\pi} \sigma(\omega) = \beta \frac{\langle \mathcal{J}_Q^2 \rangle}{L} = \frac{n}{m}$ with $n$ the density and $m = \frac{1}{2}$ the mass. This means that expansion potential is simply related to pressure $\mathcal{G}_Q = -\frac{\Omega}{Lm} = \frac{P}{m}$ with $\Omega$ the thermodynamic grand potential. The Drude thermopower is then given by $\langle \mathcal{J}_Q \mathcal{J}_E \rangle/L = \frac{T}{m}(u+P)$ with $u$ the internal energy density. We plot these quantities as a function of $\beta$ and $\mu$ for the Lieb-Liniger gas with fixed interaction strength $c=1$ by solving numerically the TBA equations (see Fig. 3). We emphasize that the relation $\mathcal{G}_Q = P/m$ between the expansion potential and pressure is general and does not rely on integrability as it follows from Galilean invariance, and also holds in higher dimensions. The only step where the integrability of the Lieb-Liniger model was crucial is to compute explicitly $P$ as a function of temperature and chemical potential or particle density.

Even though the particle expansion potential becomes especially simple in the continuum limit for systems with emergent Galilean invariance, we emphasize that its existence is more general. In particular, the expansion potential $\mathcal{G}_Q$ can also be computed exactly for the so-called $q$-Bose model [13], one of the (integrable) lattice regularizations of the Lieb-Liniger model. The $q$-Bose model has an exact conserved particle current not related to momentum, and the corresponding Drude weight can be computed using TBA equations [11]. The Drude thermopower and the expansion potential $\mathcal{G}_Q$ then follows from integrating the differential $d\mathcal{G}_Q$.

## NATURE OF THE NONEQUILIBRIUM STEADY-STATE

To investigate further the nature of the steady state and transient regions, we plot the point energy current $j_E(x,t)$ against the reduced variable $u = x/t$ in the XXZ chain at $\Delta = 0.5$ for various temperatures $T_L$, $T_R$ and $\mu = 0$. We find that the current converges to a limit shape $j_E = f(x/t)$, with $\int du f(u) = \Delta \mathcal{G}$. At low temperature, conformal field theory results imply the existence of a ballistic steady-state region with uniform current [3, 14], spreading at the sound velocity $v$, sharply separated by "shock waves" [15, 16] from the rest of the system outside the light cone. Our numerical results (Fig. 4) suggest that even at moderate temperatures, the irrelevant perturbations to the Luttinger liquid theory breaking Lorentz invariance modify that picture dramatically, reducing the steady-state region to $u = x/t = 0$ – indicating a steady-state spreading sub-ballistically. We also observe broad transient regions that propagate linearly with time. In the non-interacting case (XX spin chain), the limit shape can be expressed as $j_E = f(x/t) = g(x/t, T_L) - g(x/t, T_R)$, with

$$g(u = x/t, T) = \int_0^\pi \frac{dk}{2\pi} \frac{\epsilon_k v_k}{1 + e^{\epsilon_k/T}} \left( \theta(u + v_k) - \theta(u - v_k) \right), \quad (14)$$

with $\theta$ the Heaviside step function, $\epsilon_k = -\cos k$ and $v_k = \partial_k \epsilon_k$. The interacting case is much more involved, and it would be very interesting to investigate whether the recently developed hydrodynamic approaches for relativistic systems [15, 16] can be modified to describe such nonequilibrium steady-states in integrable spin chains.



\end{document}